%% file: EXO-12-010_temp.tex
\pdfoutput=1

\documentclass[11pt,twoside,a4paper,cmspaper,final,collab]{cms-tdr}
\def\svnVersion{183695}\def\cmsCernNo{2013-010}\def\cmsCernDate{\today}\def\cmsMessage{Submitted to Physical Review D}

\begin{document}\cmsNoteHeader{EXO-12-010}

\hyphenation{had-ron-i-za-tion}
\hyphenation{cal-or-i-me-ter}
\hyphenation{de-vices}

\RCS$Revision: 177632 $
\RCS$HeadURL: svn+ssh://svn.cern.ch/reps/tdr2/papers/EXO-12-010/trunk/EXO-12-010.tex $
\RCS$Id: EXO-12-010.tex 177632 2013-03-21 19:25:28Z hoepfner $
\newlength\cmsFigWidth
\ifthenelse{\boolean{cms@external}}{\setlength\cmsFigWidth{0.85\columnwidth}}{\setlength\cmsFigWidth{0.4\textwidth}}
\ifthenelse{\boolean{cms@external}}{\providecommand{\cmsLeft}{top}}{\providecommand{\cmsLeft}{left}}
\ifthenelse{\boolean{cms@external}}{\providecommand{\cmsRight}{bottom}}{\providecommand{\cmsRight}{right}}

\providecommand\Wprime{\PWpr\xspace}
\providecommand\Zprime{\cPZpr\xspace}
\providecommand\MT{\ensuremath{M_\mathrm{T}}\xspace}
\providecommand\MTlower{\ensuremath{M_\mathrm{T}^\mathrm{min}}\xspace}
\providecommand\ET{\ensuremath{E_\mathrm{T}}\xspace}
\providecommand\pT{\ensuremath{p_\mathrm{T}}\xspace}
\providecommand\sm{standard model\xspace}
\providecommand\invpb{\pbinv}
\providecommand\invfb{\fbinv}
\providecommand\WprimeMuNu{\ensuremath{\Wprime \rightarrow \mu\nu}\xspace}
\providecommand\WprimeENu{\ensuremath{\Wprime \rightarrow \Pe\nu}\xspace}
\providecommand\WprimeKK{\ensuremath{\cmsSymbolFace{W}_\mathrm{KK}}\xspace}
\providecommand\WprimeKKn{\ensuremath{\cmsSymbolFace{W}^n_\mathrm{KK}}\xspace}
\providecommand\WprimeKKtwo{\ensuremath{\cmsSymbolFace{W}^{n=2}_\mathrm{KK}}\xspace}
\providecommand\CI{contact interaction\xspace}

\providecommand\newb{\begin{color}{blue}}
\providecommand\newe{\end{color}}

\cmsNoteHeader{EXO-12-010} 
\title{Search for new physics in final states with a lepton and missing transverse energy in \texorpdfstring{$\Pp\Pp$}{pp} collisions at the LHC}

\date{\today}

\abstract{This letter describes the search for an enhanced production rate of events with a charged lepton and a neutrino in
high-energy $\Pp\Pp$ collisions at the LHC. The analysis uses data collected with the CMS detector, with an integrated
luminosity of 5.0\invfb at $\sqrt{s} = 7\TeV$, and a further 3.7\invfb at $\sqrt{s} = 8\TeV$.
No evidence is found for an excess.
The results are interpreted in terms of limits on a heavy charged gauge boson (\Wprime) in the sequential standard model, a split
universal extra dimension model, and contact interactions in the helicity-non-conserving model. For the last, values of the binding energy below 10.5
(8.8)\TeV in the electron (muon) channel
are excluded at a 95\% confidence level. Interpreting the  $\ell\nu$ final state in terms of a heavy
\Wprime with standard model couplings, masses below
2.90\TeV are excluded.}

\hypersetup{%
pdfauthor={CMS Collaboration},%
pdftitle={Search for new physics in final states with a lepton and missing transverse energy in pp collisions at the LHC},%
pdfsubject={CMS},%
pdfkeywords={CMS, physics, W' decay, split UED, contact interaction}}

\maketitle 

\section{Introduction}

New heavy gauge bosons  are predicted by various
extensions of the standard model (SM).
The sequential standard model (SSM)~\cite{reference-model} postulates
the existence of a \Wprime boson,
a heavy analogue of the $\PW$. In such a theory, the \Wprime is expected to appear as a narrow
resonance with decay modes and branching fractions similar
to those of the $\PW$. For \Wprime masses above 180\GeV, where the $\cPqt\cPaqb$ decay channel opens up,
the predicted branching fraction is about 8.5\% for each of the leptonic final states.
Previous searches~\cite{cms-2011limit, atlas2011} with $\Pp\Pp$ collision data at $\sqrt{s} = 7\TeV$
by the Compact Muon Solenoid (CMS) and ATLAS experiments at the Large Hadron Collider (LHC),
based on an integrated luminosity of up to 5\invfb,
have excluded SSM \Wprime bosons with masses up to 2.6\TeV.

Other models for new physics predict the same final state, such as those with universal extra dimensions (UED) and bulk fermions, or
split UED~\cite{PhysRevD.79.091702, JHEP04(2010)081}.
Such models of extended space-time assume one additional compact
dimension of radius $R$.
The split-UED parameter space is defined by $1/R$ and $\mu$, with $\mu$ being the bulk mass parameter of the fermion field in five dimensions.
For suitable non-zero values of $\mu$, as assumed by split-UED models,
the cross sections are sufficiently large to allow observation by LHC experiments.
All SM particles have corresponding
Kaluza-Klein (KK) partners, for instance \WprimeKKn,
where $n$ denotes the $n$-th KK excitation mode.
Only KK-even modes of \WprimeKKn couple to SM fermions, owing to
KK-parity conservation~\cite{PhysRevD.64.035002}.
Modes with $n \geq 4$ have a smaller cross section and are not expected to be accessible at $\sqrt{s} = 8\TeV$, hence
the only mode considered is $n = 2$.

Motivated by the observation of mass hierarchies in the fermion
sector, theories have been developed where leptons and quarks are composite objects~\cite{preons}.
At energies much lower than the binding energy of these fundamental constituents, typically called $\Lambda$,
quark and lepton compositeness
would manifest itself as a four-fermion contact interaction (CI). One of the possible
contact interactions between two quarks, a
neutrino and a charged lepton is described by the helicity-non-conserving (HNC) model~\cite{HNCM-CI}.
The corresponding cross section is
proportional to the square of the partonic centre-of-mass energy and to $\Lambda^{-4}$.
While CDF has set a limit on $\Lambda> 2.81\TeV$ ~\cite{cdf-CIlimit}
based on a final state with an electron and a neutrino,
no limit in the HNC model has yet been set in the muon channel.

In this letter, a search is presented for an excess of events with an isolated charged lepton (an electron or muon) and a neutrino in the final state,
using the CMS detector. The data sample corresponds to an integrated luminosity of 5.0\invfb
at $\sqrt{s} = 7\TeV$ collected in 2011, and
3.7\invfb at $\sqrt{s} = 8\TeV$ collected in 2012. The CMS 2011 result has been published in ref.~\cite{cms-2011limit}.

\section{The CMS detector}
The central feature of the CMS apparatus is a superconducting solenoid of 6\unit{m} internal
diameter, providing a magnetic field of 3.8\unit{T}. Within the superconducting solenoid volume are a silicon pixel and
strip tracker, a lead tungstate crystal electromagnetic calorimeter (ECAL), and a brass/scintillator hadron calorimeter
(HCAL). Muons are measured in gas-ionization detectors embedded in the steel return yoke. Extensive forward calorimetry
complements the coverage provided by the barrel and endcap detectors.
A more detailed description of the CMS detector can be found in Ref.~\cite{JINST}.
CMS uses a right-handed coordinate system, with the origin at the nominal interaction point, the $x$ axis pointing to
the centre of the LHC, the $y$ axis pointing up (perpendicular to the LHC plane), and the $z$ axis along the
anticlockwise-beam direction. The polar angle $\theta$ is measured from the positive $z$ axis and the azimuthal angle
$\phi$ is measured in the $x$-$y$ plane.

\section{The search strategy}
Candidate events with at least one high transverse momentum (\pT)
electron or muon are selected using single-lepton
triggers.
Isolated high-\pT leptons are reconstructed using very stringent quality criteria while the neutrino gives rise to experimentally observed missing transverse energy (\MET).
The details on lepton identification and \MET reconstruction, as optimized for
the 2011 analysis, can be found in Ref.~\cite{cms-2011limit}.
The discriminating variable of this analysis is the transverse mass \MT of the lepton--\MET system,
calculated as
\begin{align}
\MT = \sqrt{2 \cdot \pt^\ell  \cdot \MET \cdot (1 - \cos \Delta
  \phi_{\ell,\nu}) },
\end{align}
where $\Delta \phi_{\ell,\nu}$ is the azimuthal angle between
the charged lepton's transverse
momentum $\pt^\ell$ and the \MET direction.

\begin{figure}[hbtp]
\begin{center}
 \includegraphics[width=0.47\textwidth]{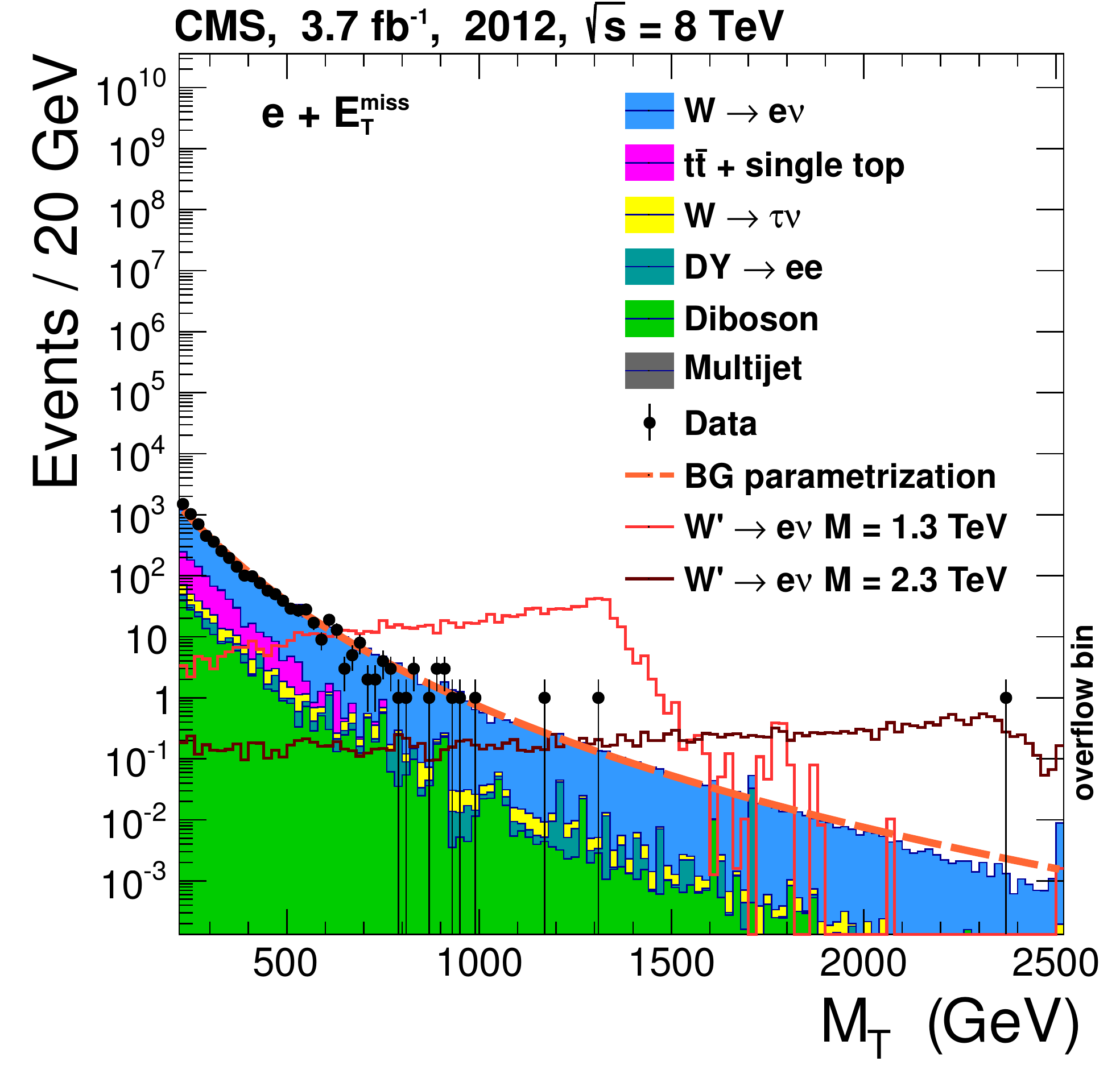}
 \includegraphics[width=0.47\textwidth]{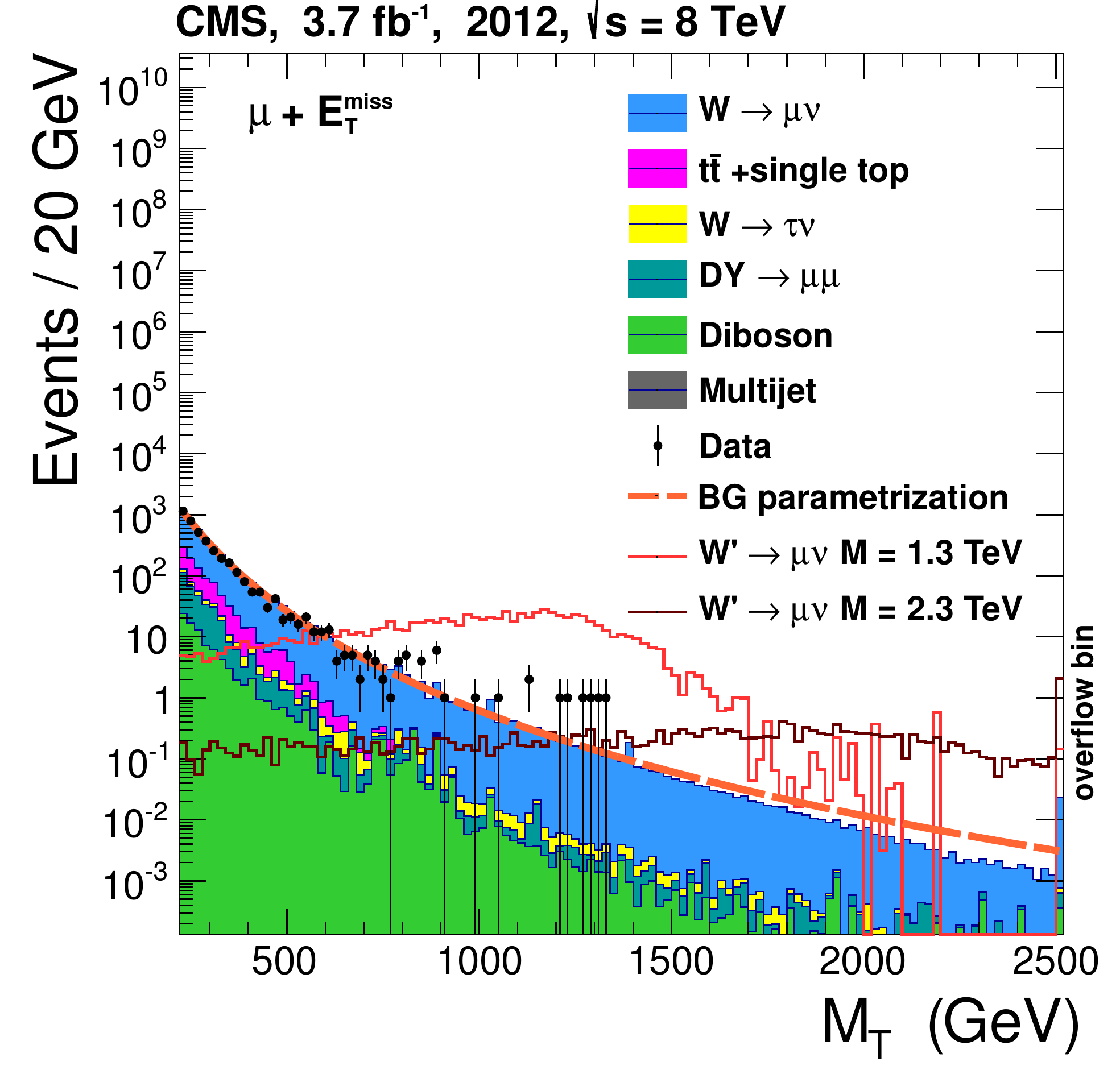}
\caption{Observed lepton-\MET transverse mass distributions in the electron
 (\cmsLeft) and the muon (\cmsRight) channel.
The dashed lines show the parametrization of the background as described in the text
(labelled as BG paramatrization
in the legend).
Simulated signal distributions for a SSM \Wprime are also
shown, including detector resolution effects.
The simulated background labeled as `diboson' includes $\PW\PW$,
 $\cPZ\cPZ$, and $\PW\cPZ$ contributions. The last bin contains the contributions of all bins above the displayed range.
}
  \label{fig:MT}
\end{center}
\end{figure}

\begin{figure}[hbtp]
\begin{center}
 \includegraphics[width=0.49\textwidth]{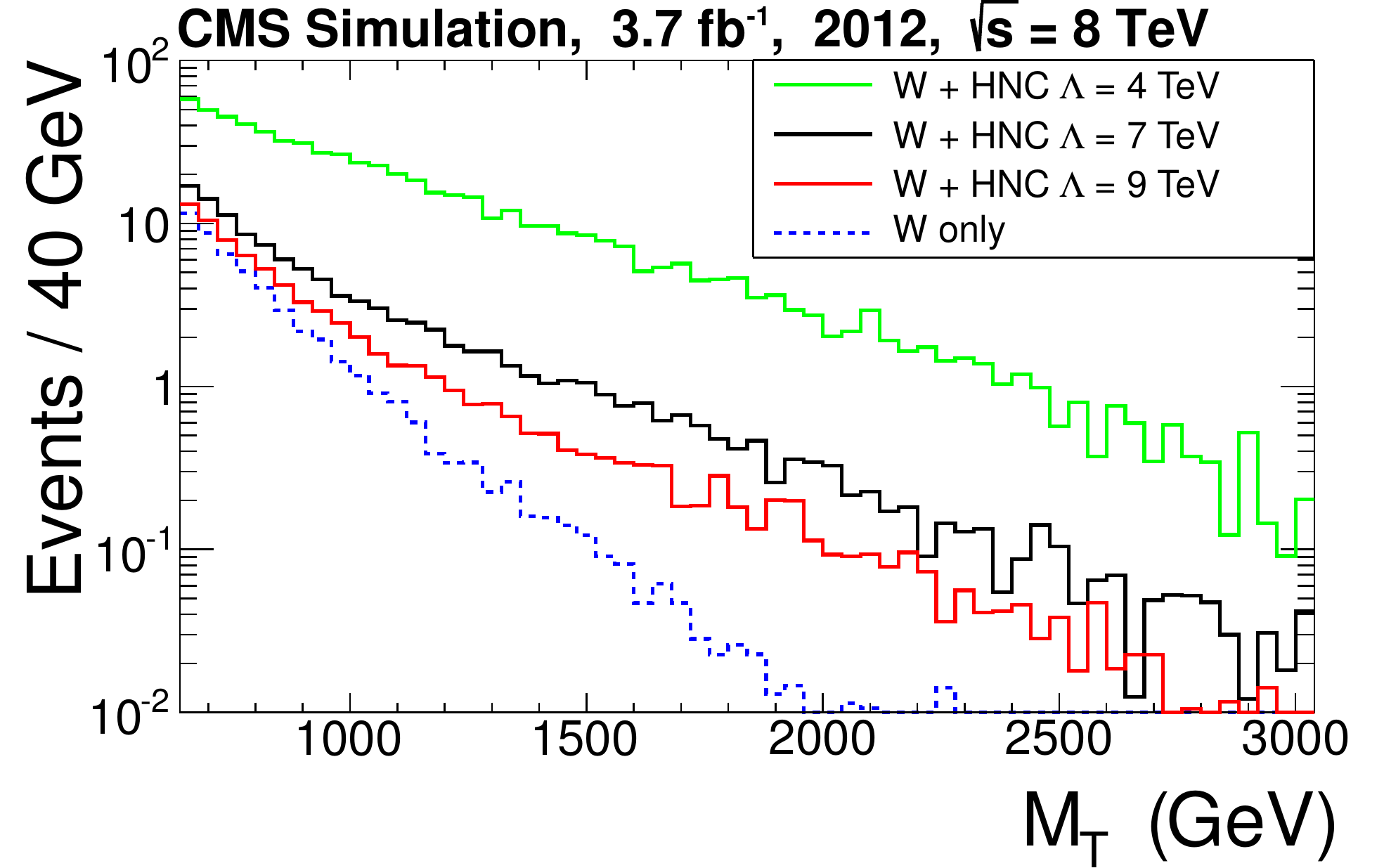}
 \caption{Simulated transverse mass distribution shown for contact interaction (HNC model),
 in the muon+\MET channel at generator level for CI signals with $\Lambda$ = 4, 7, and 9\TeV,
 and the SM background ($\PW\rightarrow \Pgm \Pgn$).
}
\label{fig:CIsignal}
\end{center}
\end{figure}

In \Wprime decays, as well as for the other models considered,
the lepton and \ETmiss are expected to be almost back-to-back in the transverse plane,
and balanced in transverse energy. Additional kinematic criteria therefore
select events with a ratio of the lepton $\pt^\ell$ and
the \ETmiss, $0.4 < \pt^\ell/{\ETmiss} < 1.5$, along with the requirement
of the angular difference,
$|\Delta \phi_{\ell,\nu} - \pi| < 0.2 \pi$.
For simulated SSM \Wprime events with masses between 0.5 and 2.5\TeV passing these selection criteria, the signal efficiency
(including 90\% geometrical acceptance) is found to be
70--75\% with 2\% uncertainty in the electron channel and
67--72\% with 1\% uncertainty in the muon channel.
For the HNC contact-interaction model the signal efficiency is independent
of the interaction scale $\Lambda$ and
has been determined from simulation to be 80\% with 1\% uncertainty
for the electron channel and
77\% with 4\% uncertainty for the muon channel.
The transverse mass distributions for accepted SM events in the electron and muon channels are shown in
Fig.~\ref{fig:MT}, along with two example \Wprime signals.
The observed event with the highest transverse mass in the electron channel has \MT{} = $2.38\pm0.05$\TeV based on
$\ET=1.2\TeV$ with 1\% uncertainty.
In the muon channel the maximum transverse mass is $\MT{} = 1.33\pm0.03\TeV$ with a measured $\pt^\ell$ of $690\pm22\GeV$.
The two types of processes, \Wprime production and compositeness, can be distinguished by
examining the shape of the lepton-\MET transverse-mass spectrum. Both processes manifest themselves through
an excess of events at the high end of the spectrum.
The \Wprime signal events are expected to concentrate in a Jacobian peak around the \Wprime mass,
as shown in Fig.~\ref{fig:MT}.
Compositeness would rather yield an unstructured excess, with the excess relative to the standard model contribution
increasing as a function of \MT, as shown in Fig.~\ref{fig:CIsignal}.

\section{Signal and backgrounds}

The \Wprime (this also includes \WprimeKKtwo)
and CI signals are generated at leading order (LO) with \PYTHIA~6.4.26.~\cite{Sjostrand:2006za}
using the CTEQ6L1~\cite{CTEQ6L1} parton distribution
functions (PDF).
They are scaled to the next-to-next-to-leading order (NNLO)
cross section calculated with $\textsc{fewz}$~\cite{fewz,
Gavin:2012kw} for each \Wprime (\WprimeKKtwo) mass point. In the absence of higher-order calculations,
LO cross sections are used for the CI model.

The primary source of background is the off-peak, high-\MT
tail of the standard model $\PW \to \ell\nu$ decays. Other important
backgrounds arise from \ttbar, Drell--Yan,
and diboson ($\PW\PW$, $\PW\cPZ$, $\cPZ\cPZ$) events.
An important background also comes from QCD multijet processes, but this is largely removed by the analysis selections
and therefore cannot be readily distinguished in Fig.~\ref{fig:MT}.
Contributions from decays
with $\Pgt$ leptons in the final state
that subsequently decay to an electron
or muon and neutrinos are considered as well, but found to be negligible.
The \PYTHIA~6 generator is used for all background processes
except for \ttbar, which is generated with \MADGRAPH~4~\cite{madgraph}
in combination with \PYTHIA.
The numbers of Monte Carlo events
are normalized using the integrated luminosity of the recorded data and
with NNLO cross sections, except
diboson and multijet samples, for which the next-to-leading order and
LO cross sections are used, respectively.
In all event samples, additional minimum bias interactions are superimposed onto the main background processes to match the luminosity profile of the analyzed data set.

The background prediction is based on the transverse-mass distributions of these
simulations, as seen in Fig.~\ref{fig:MT}. The summed background distribution is
parametrized
with a function of form
$	f(\MT) = \frac{a}{(\MT + b)^c} $
to avoid event fluctuations due to the limited size of the Monte Carlo event sample,
also shown in Fig.~\ref{fig:MT}.
In the low mass control region, e.g. for \MT $>$ 500\GeV,
the background prediction in the electron
channel is 169 events from a direct count and 155$\pm$16 events from the parametrization. The
corresponding event numbers in the muon channel are 162 and 138$\pm$12, from the direct
count and the parametrization respectively.
The number of observed events in each channel is 153.
The expected background due to
SM processes drops significantly with increasing \MT.
The number of events
observed and the number expected from the parametrization, above three characteristic
high
\MT thresholds, are shown in Table~\ref{tab:exampleEventYields}
for the electron and the muon channels.
Also summarised are
the expected numbers of \Wprime and
CI signal events, as evaluated from simulation,
including systematic uncertainties due to imperfections in the description of the
detector performance.

\begin{table}
 \topcaption{Data, background, and signal event yields for different transverse mass thresholds.}
\label{tab:exampleEventYields}
\centering
\renewcommand{\arraystretch}{1.4}
\begin{scotch}{lccc}
& $\MT>1.0\TeV$& $\MT>1.5\TeV$& $\MT>2.0\TeV$\\ \hline
\multicolumn{4}{c}{Electron channel}\\\hline
Data                                     & 3                          & 1                         &1\\
SM background                            & $6.8\pm0.4$                & $0.69^{+0.05}_{-0.04}$    &$0.13\pm0.02$\\
$M_\Wprime=2.5 \TeV$         & $9.7\pm0.3$                & $7.5\pm0.2$               &$4.7\pm0.2$ \\
$M_\Wprime=3 \TeV$           & $1.98\pm0.06$              & $1.5\pm0.05$              &$1.15^{+0.05}_{-0.04}$ \\
CI, $\Lambda=4 \TeV$ & $220\pm5$                  & $72\pm3$                  &$21\pm1$ \\
CI, $\Lambda=9 \TeV$ & $8.6\pm0.2$                & $2.8\pm0.1$               &$0.81^{+0.05}_{-0.04}$ \\
\hline
\multicolumn{4}{c}{Muon channel}\\\hline
Data& 9& 0&0\\
SM background& $6.5^{+1.9}_{-1.5}$& $0.92^{+0.49}_{-0.32}$         &$0.20^{+0.14}_{-0.08}$\\

$M_\Wprime=2.5 \TeV$& $9.4\pm0.3$& $7.0\pm0.5$               &$4.1^{+0.8}_{-0.7}$\\
$M_\Wprime=3 \TeV$ &$1.9\pm0.1$& $1.4\pm0.1$                 &$0.94^{+0.12}_{-0.10}$\\
CI, $\Lambda=4 \TeV$ & $240\pm 20$    & $80^{+15}_{-13}$   &$28^{+8}_{-6}$ \\
CI, $\Lambda=9 \TeV$ & $34\pm 3$      & $12^{+3}_{-2}$    &$3.5^{+1.2}_{-0.9}$ \\

\end{scotch}
\end{table}

\begin{figure}[hbtp]
\begin{center}
 \includegraphics[width=0.49\textwidth]{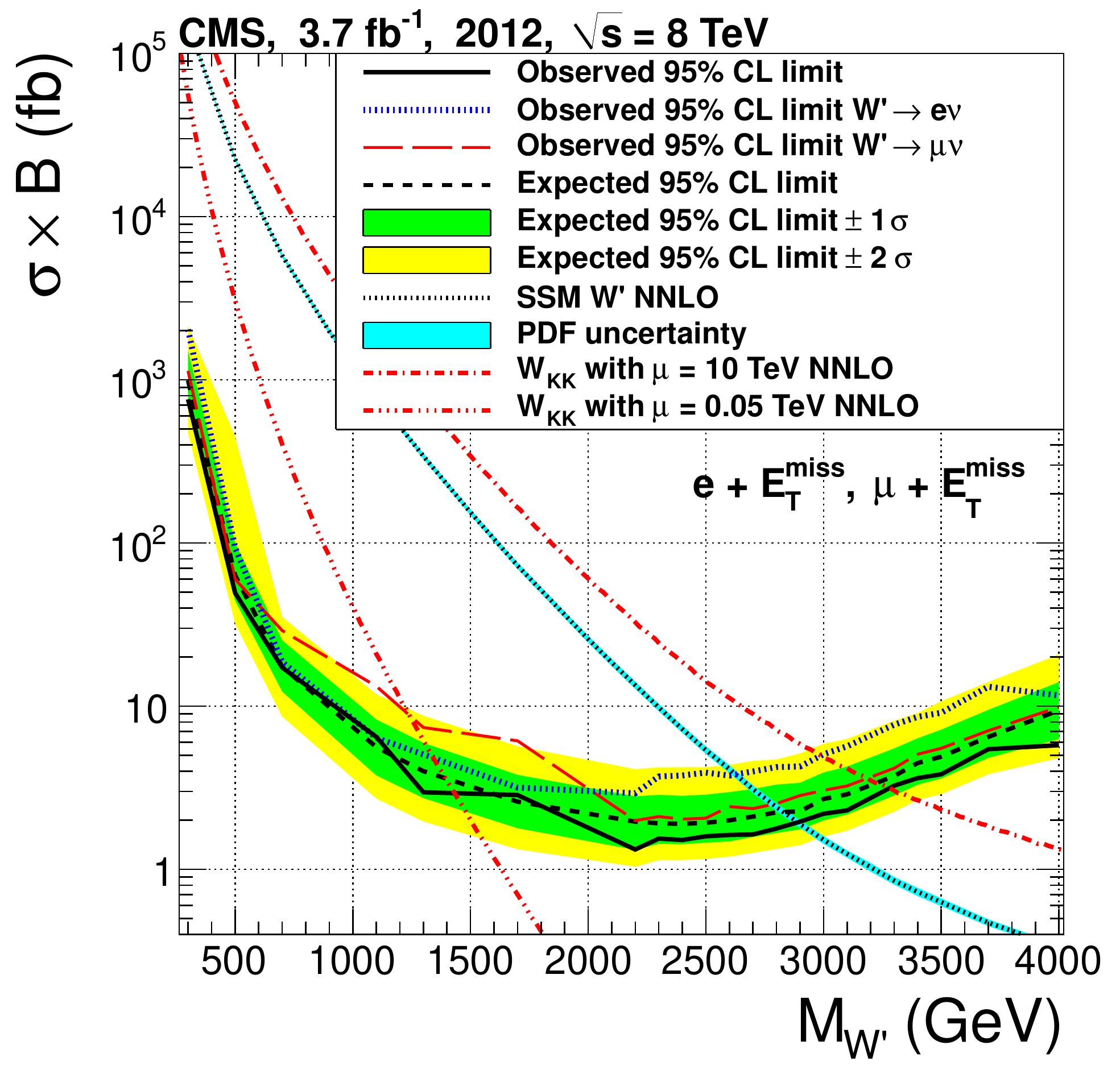}
 \caption{Limits on the cross section times the single channel branching fraction
($\sigma \times \mathcal{B}$)
for heavy \Wprime bosons
based on the 2012 data for the electron and the muon channels.
For the individual channels, only the observed limits are shown. For the combination, the observed limit, the expected limit,
the 1$\sigma$, and the 2$\sigma$ bands are displayed.
The model assumes equal branching fractions for the electron and the muon
channel, hence the combination corresponds to doubling the number of events.
All limits are displayed for the single channel branching fraction.
The \Wprime mass limits are derived with a
Bayesian method
for the models of a SSM \Wprime and \WprimeKKtwo in split UED.
The \WprimeKKtwo is the lowest mass state that can
couple to SM fermions and has the same final state as the
SM-like \Wprime. Since it has even KK parity it can be produced singly.
}
\label{fig:CombLimit2012}
\end{center}
\end{figure}

\begin{figure*}[hbtp]
\begin{center}
 \includegraphics[width=0.32\textwidth]{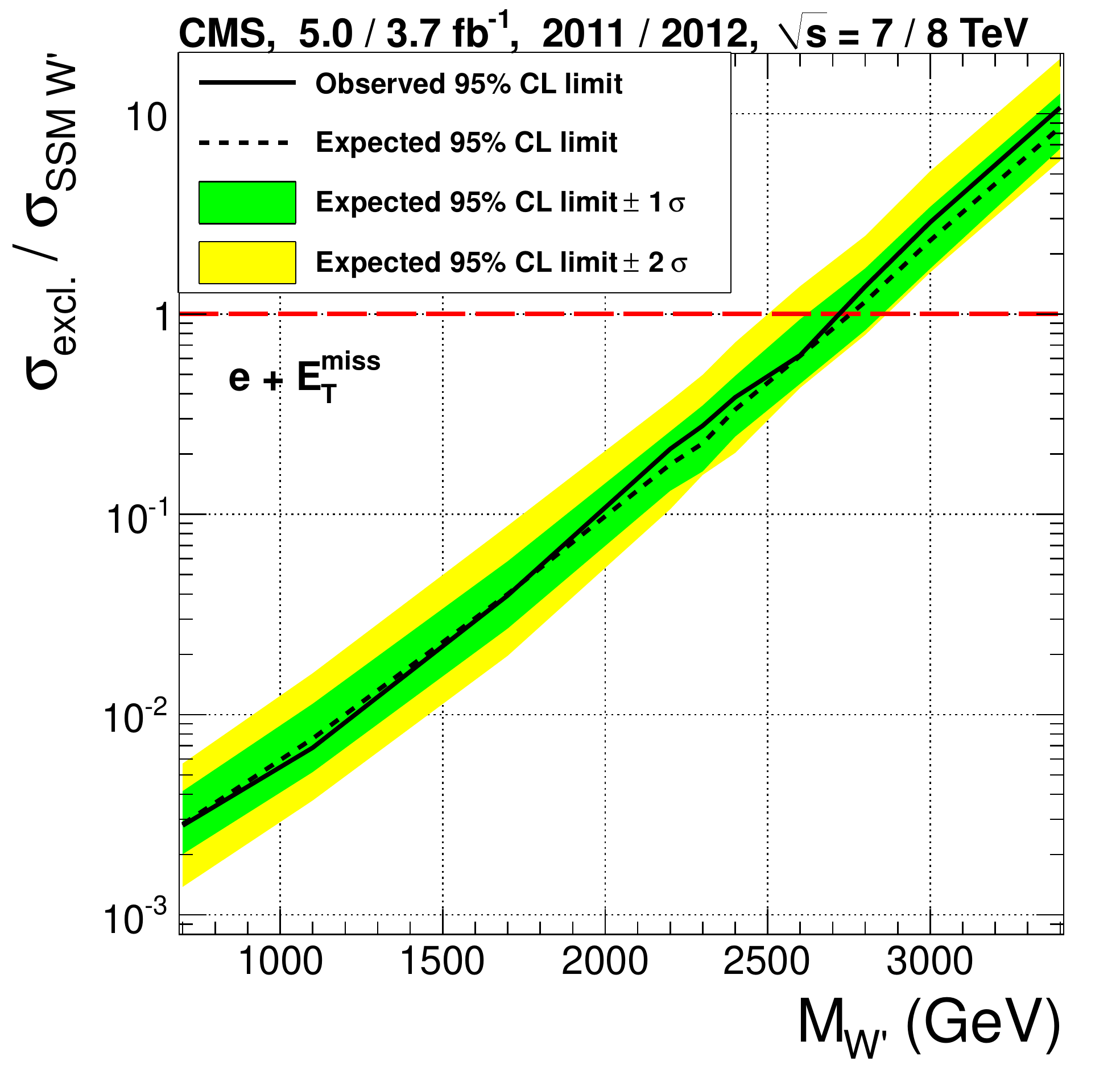}
 \includegraphics[width=0.32\textwidth]{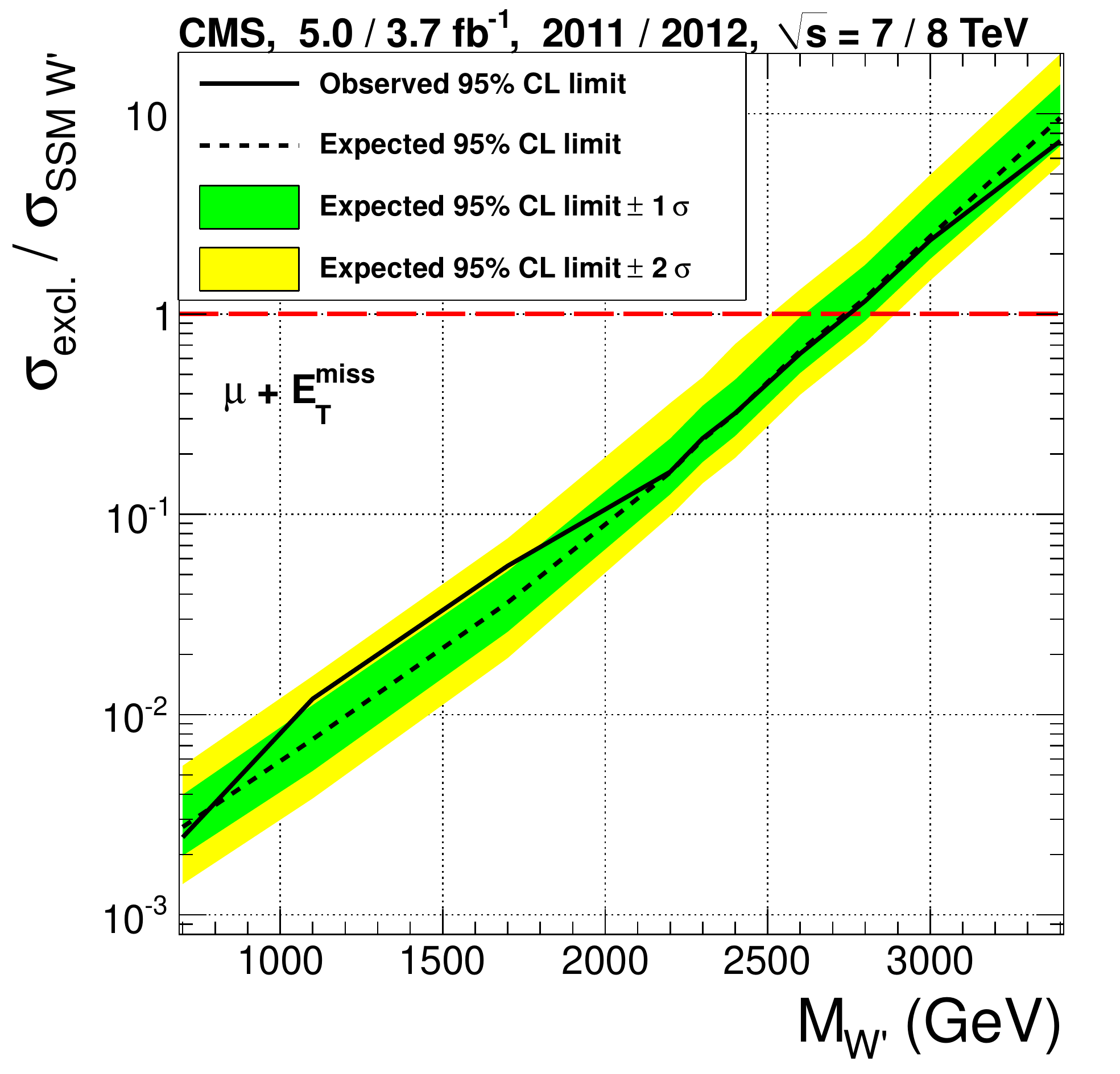}
 \includegraphics[width=0.32\textwidth]{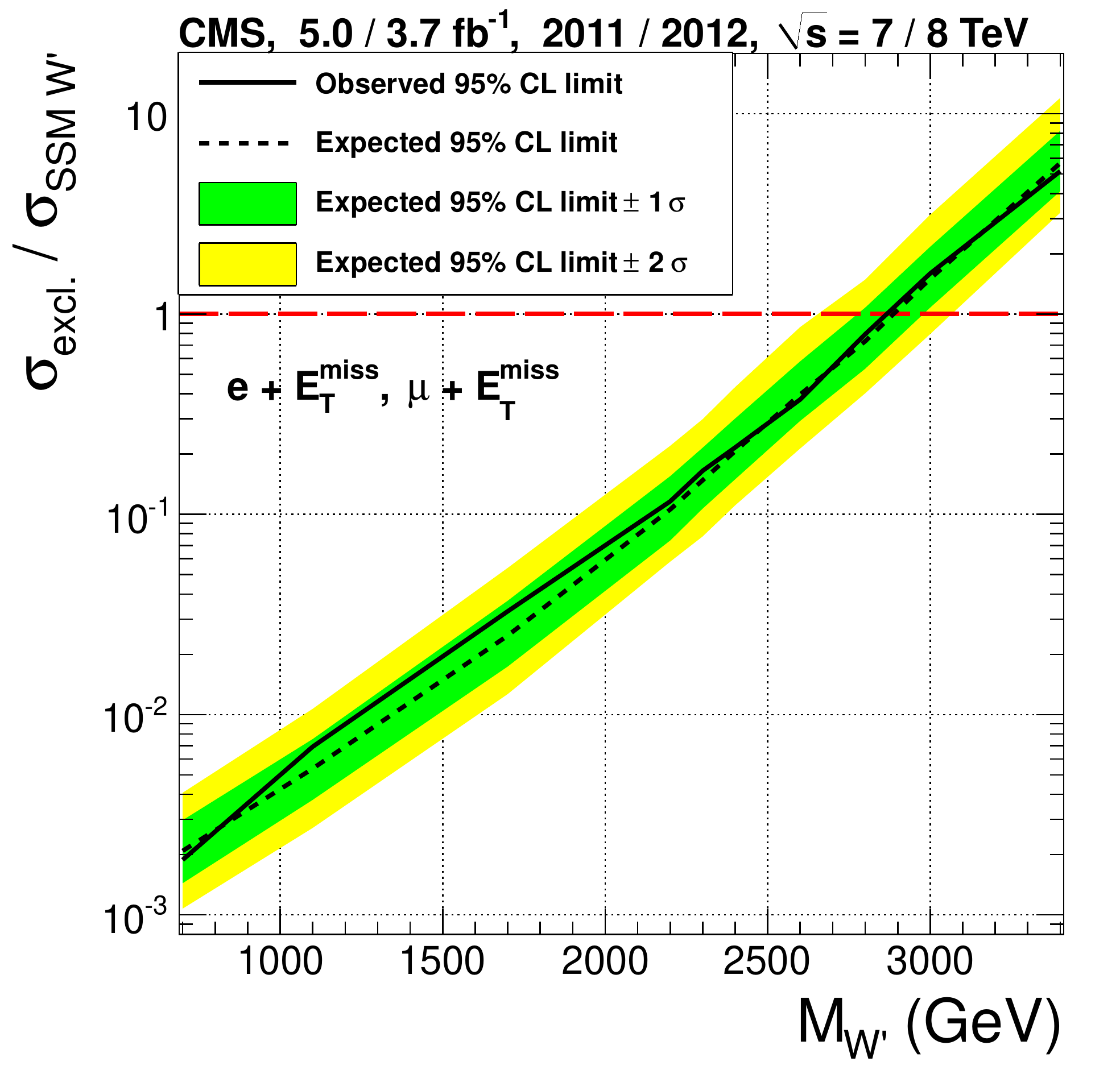}
 \caption{Combination of 2011 and 2012 data for the electron and the muon channels using
5.0\invfb of data at $\sqrt{s}=7\TeV$ from 2011 and 3.7\invfb of data at $\sqrt{s}=8\TeV$ from 2012.
The upper row shows the individual combinations for the electron (left) and muon channel (right). 
To the right the combination of both is displayed.
The limits are derived with a Bayesian method.
Plotted is the signal strength modifier
$\sigma_\text{excl.}/\sigma_\text{SSM \Wprime}$ as a function of the \Wprime mass,
where $\sigma_\text{excl.}$ is the cross section excluded at a 95\% confidence level and $\sigma_\text{SSM \Wprime}$
is the cross section predicted by the SSM.
All \Wprime mass points below the ratio $\sigma_\text{excl.}/\sigma_\text{SSM \Wprime} = 1$, shown as a red dashed line, are excluded in the sequential standard model.
}
\label{fig:CombLimit20112012}
\end{center}
\end{figure*}

\begin{figure}[hbtp]
\begin{center}
 \includegraphics[width=0.49\textwidth]{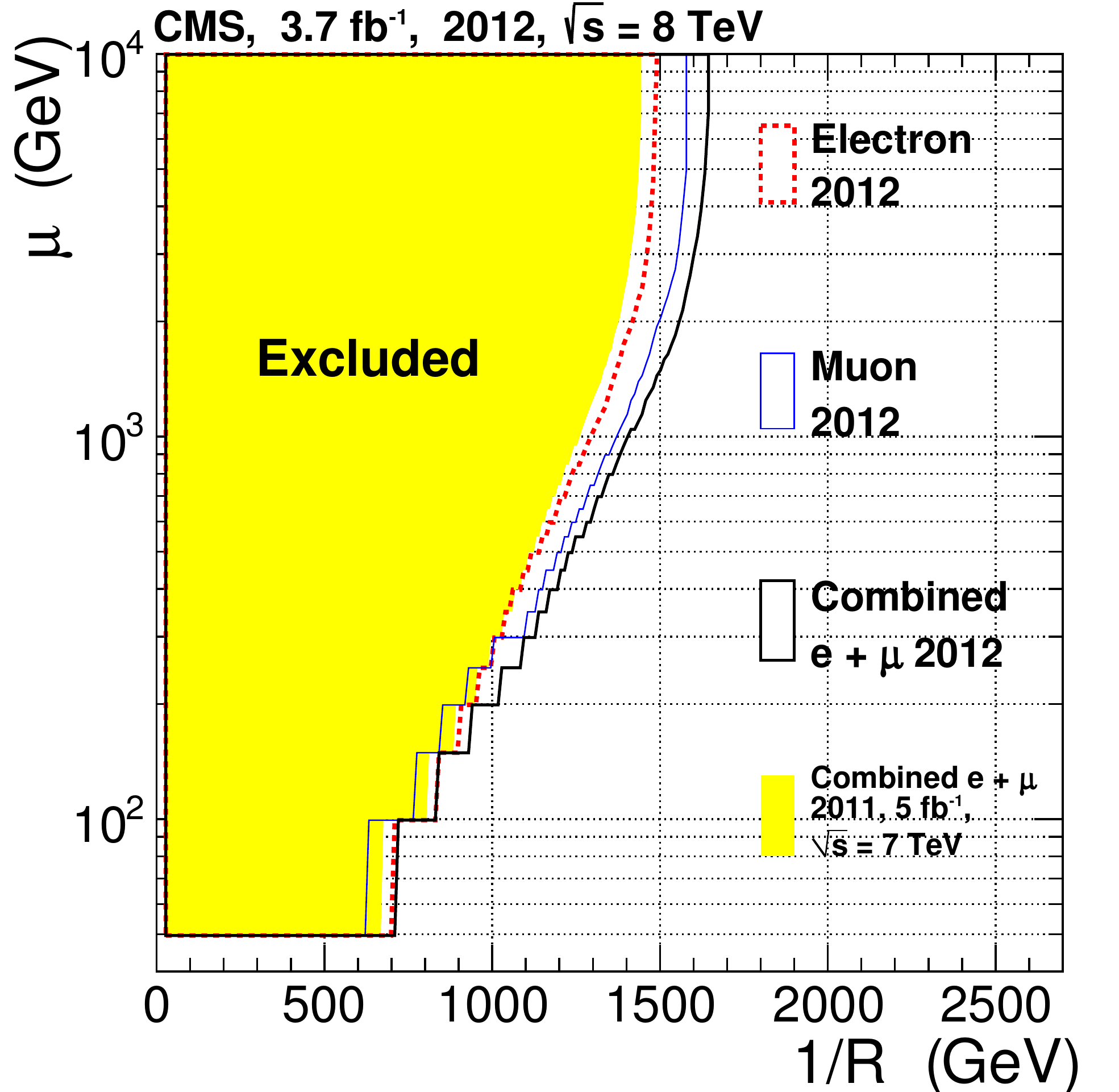}
 \caption{The 95\% confidence limits on the split-UED parameters $\mu$ and $1/R$
derived from the \Wprime mass limits taking into account the
corresponding width of the \WprimeKKtwo. For the 3.7\invfb of 2012 data the individual
limits for the electron and muon channels are shown together with their combination,
improving
the excluded parameter space based on the 2011 data shown in yellow.
}
\label{fig:splitUED2012}
\end{center}
\end{figure}

\section{Results and exclusion limits}

For determining the exclusion limits, systematic uncertainties on the signal and background yields
are included via
nuisance parameters with a log-normal prior distribution
in the same way as in our previous analysis from Ref.~\cite{cms-2011limit}.
The lepton-energy calibration and resolution distort the transverse mass spectrum.
Uncertainties in the energy scale and the energy resolution make comparable contributions to the
overall uncertainty in the electron channel: the energy scale has an uncertainty of 1(3)\% in the
barrel(endcaps) and the resolution has an uncertainty of 1.4(3)\% for the barrel
(endcaps)~\cite{DP-12-007}.
For muons, the dominant uncertainty stems from the momentum scale, which is taken to be
5\% $\times$ \pT / \TeV~\cite{muo-11-001}
and results in larger errors on the background and signal event yields in the muon
channel, as can be seen in Table~\ref{tab:exampleEventYields}.
The muon \pT resolution has been determined with cosmic ray muons to be 10\% at high
\pT with an uncertainty of 0.6\%~\cite{muo-11-001}.

Similarly, the impact of the \MET energy scale is modelled
by shifting the hadronic component event-by-event by
10\%, while the resolution is taken into account by a 10\% smearing.
In all cases, the impact on the expected number of signal events is around 1\%
for each source of uncertainty.
The background parametrization procedure is repeated using the distorted
distributions.
Additionally, an uncertainty is derived by fitting
the undistorted background with two different functions.
The estimates shown in Table~\ref{tab:exampleEventYields}
include a systematic
uncertainty that covers the range of results from these fits, with the
statistical uncertainty coming from the fit to the original
distribution.
Additionally, an uncertainty of 4.4\% is considered on the integrated luminosity~\cite{LUM-12-001}.

The number of data events above a transverse mass
threshold \MTlower is compared to the expected number of signal and
background events, with the
\MTlower threshold being optimized for the best expected exclusion limit,
using steps of 50\GeV.
Very similar results are achieved when optimizing for discovery.
The same optimization procedure is applied for the \Wprime and
CI searches but the optimized \MT thresholds are specific to each
analysis.
For a \Wprime mass of up to 2.5\TeV
the optimization tends to select a value for \MTlower a little below
   the peak of the Jacobian, retaining most of the events in the peak.
To keep a significant signal contribution for higher masses, the \MTlower threshold is shifted to lower values
owing to the increasing off-shell fraction.
For CI no such effect exists and
the optimized \MTlower threshold is roughly constant around 1\TeV.

No significant excess has been observed in the data, and upper limits are set
on the production cross section times the
branching fraction
$\sigma \times \mathcal{B}(\Wprime \to \ell \cPgn)$, with $\ell
= \Pe$ or $\mu$, using a Bayesian method with the assumption of a flat prior for the parameter of interest.
The expected and observed upper cross section limits on a SSM \Wprime, at 95\% confidence level (CL), for $\sqrt{s}$ = 8\TeV data,
are shown in Fig.~\ref{fig:CombLimit2012} for both channels and their combination.
Using the central value of the theoretical NNLO SSM \Wprime cross section times
branching fraction, which is assumed to be identical in the two channels, we exclude
masses less than
2.60\TeV in the electron and
2.75\TeV in the muon channel.
The expected limits are 2.70\TeV and 2.65\TeV, respectively.
The observed limit for \WprimeENu is slightly
lower than the expectation because of one event with a transverse mass of 2.3\TeV (Fig.~\ref{fig:MT}).
Combining both channels, the limit increases to 2.85\TeV (Table~\ref{tab:AllLimits20112012}).
To further enhance the sensitivity the new results can be combined
with the published~\cite{cms-2011limit} $\sqrt{s}$ = 7\TeV results, based on an integrated luminosity of 5~\invfb, thus extending
the limit to 2.90\TeV as shown in Fig.~\ref{fig:CombLimit20112012} and Table~\ref{tab:AllLimits20112012}.

\begin{figure}[h!bt]
\begin{center}
 \includegraphics[width=0.49\textwidth]{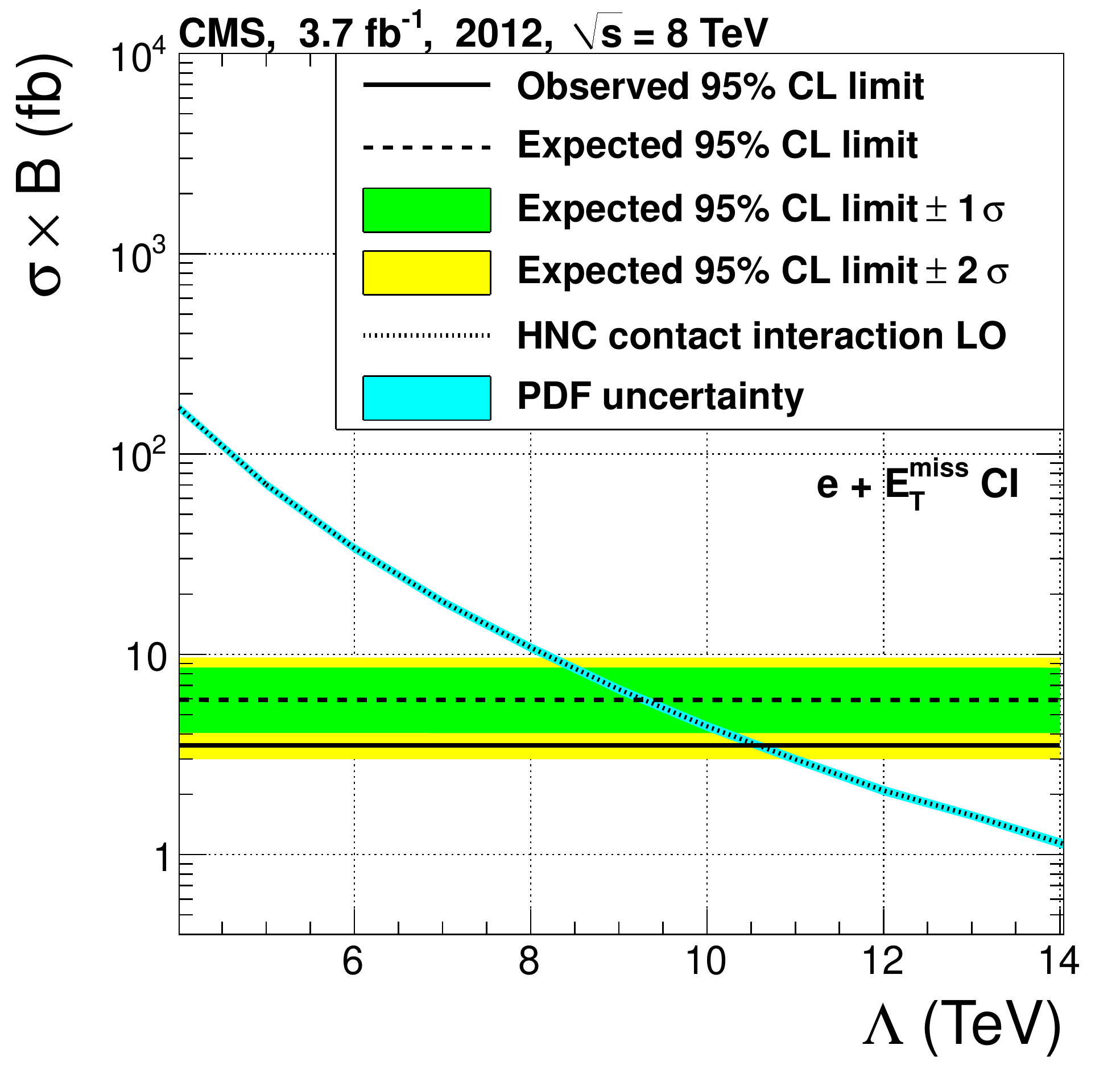}
 \includegraphics[width=0.49\textwidth]{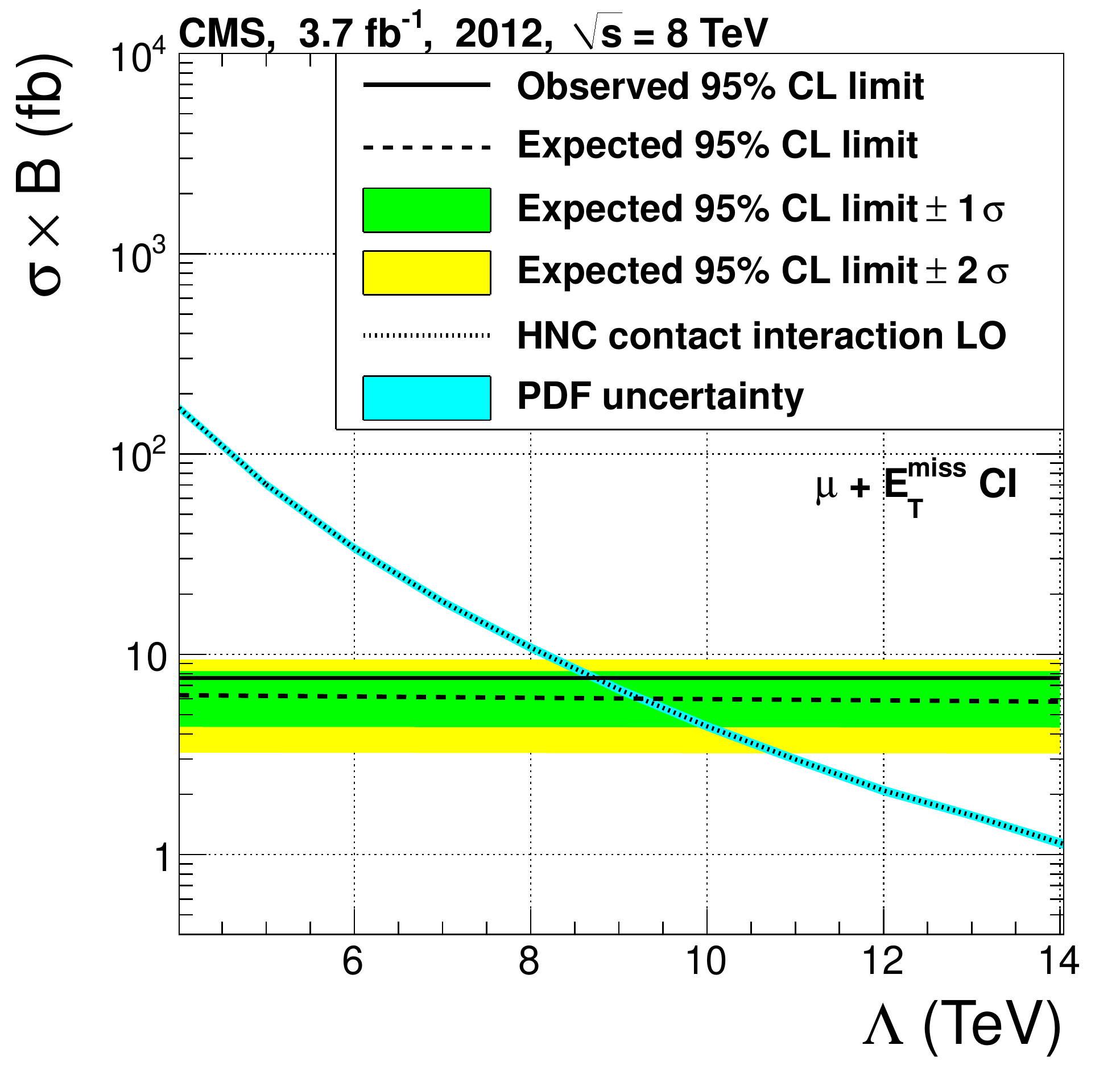}
 \caption{Bayesian limits
on the cross section times the single channel branching fraction ($\sigma \times \mathcal{B}$)
for
contact interactions (HNC model) in the electron+\MET (\cmsLeft) and the muon+\MET channel (\cmsRight)
based on an integrated luminosity of 3.7\invfb of $\sqrt{s}$ = 8\TeV data. The expected limit in either channel
excludes values of the new interaction scale $\Lambda <$9\TeV. The observed limit, driven by the data,
excludes $\Lambda <10.5\TeV$ (8.8) in the electron (muon) channel. The signal efficiency is independent of $\Lambda$.
}
\label{fig:CILimit}
\end{center}
\end{figure}

The limit can be reinterpreted in terms of \WprimeKKtwo mass, as shown in Fig.~\ref{fig:CombLimit2012},
for values of the Dirac mass term $\mu = 0.05\TeV$ and $\mu = 10\TeV$ and
directly
translated to bounds on the split-UED parameter space, $(1/R, \mu)$
(Fig.~\ref{fig:splitUED2012}).
The four-fermion contact interaction of the HNC model
is excluded for values of
$\Lambda < 10.5\TeV$ in the electron channel and $\Lambda < 8.8\TeV$ in the muon channel
(Fig.~\ref{fig:CILimit}).
While the expected sensitivity is comparable in both search channels,
the exclusion limit in the electron channel is higher because of a downward fluctuation of data
(see Table~\ref{tab:exampleEventYields}) in the search region of $\MT>1\TeV$.
These limits are summarized in Table~\ref{tab:AllLimits}.

\section{Summary}

A search for new physics has been carried out using $\Pp\Pp$ collisions recorded with the
CMS detector at centre-of-mass energies of $\sqrt{s} = 7\TeV$ and $8\TeV$. The transverse mass spectrum of either
electrons or muons and missing transverse energy has been measured. The observed spectra are consistent
with the standard model expectation, and mass limits
on sequential standard model \Wprime and
\WprimeKKtwo have been set
at the 95\% CL.
These are the most stringent limits to date.
For the first time, a limit on the compositeness scale $\Lambda$ has been
set for contact interactions in the HNC model based on the final state with a muon and \MET.
The corresponding exclusion limit in the electron channel significantly improves the
existing one.

\begin{table}[htb]
\centering
\topcaption{Exclusion limits in \TeV on the SSM \Wprime mass for the electron and muon channel as well as their combinations
based on 5.0\invfb of 2011 data at $\sqrt{s} = 7\TeV$ and 3.7\invfb of 2012 data at $\sqrt{s} = 8\TeV.$}
\label{tab:AllLimits20112012}
\renewcommand{\arraystretch}{1.4}
\begin{scotch}{c|cc|cc|cc}

Channel & Obs. & Exp.&Obs.& Exp.&Obs.& Exp.\\
 & 2011 & 2011 &2012&2012&2011+2012&2011+2012\\
\hline
\Pe     & 2.40 &2.45 & 2.60  &  2.70   & 2.70  & 2.75  \\
 \Pgm  & 2.40 &2.45 & 2.75  &  2.65   & 2.75  & 2.70  \\
$\Pe+ \Pgm$ & 2.50 &2.60 & 2.85  &   2.80   & 2.90  & 2.90  \\

\end{scotch}
\end{table}

\begin{table}[htb]
\centering
\topcaption{Exclusion limits for other model interpretations
for the electron and muon channel or their combination
based on 3.7\invfb of 2012 data at $\sqrt{s} = 8\TeV.$}
\label{tab:AllLimits}
\renewcommand{\arraystretch}{1.3}
\begin{scotch}{l|c|c|c}

Model & Channel & Observed limit & Expected limit \\
\hline
\WprimeKKtwo & e+$\mu$ & $\rm{m_{\WprimeKKtwo}} < 1.5\TeV$& $\rm{m_{\WprimeKKtwo}} < 1.4\TeV$\\
$\mu$ = 0.05\TeV &     &       & \\ \hline
\WprimeKKtwo & e+$\mu$ & $\rm{m_{\WprimeKKtwo}} < 3.3\TeV$& $\rm{m_{\WprimeKKtwo}} < 3.2\TeV$\\
$\mu$ = 10.0\TeV &     &       & \\ \hline\hline
HNC CI & \Pe      & $\Lambda < 10.5\TeV$ &$\Lambda < 9.0\TeV$ \\
HNC CI & $\mu$  & $\Lambda <  8.8\TeV$ & $\Lambda < 9.0\TeV$ \\

\end{scotch}
\end{table}

\section*{Acknowledgements}

We congratulate our colleagues in the CERN accelerator departments for the excellent performance of the LHC and thank
the technical and administrative staffs at CERN and at other CMS institutes for their contributions to the success of
the CMS effort. In addition, we gratefully acknowledge the computing centres and personnel of the Worldwide LHC
Computing Grid for delivering so effectively the computing infrastructure essential to our analyses. Finally, we
acknowledge the enduring support for the construction and operation of the LHC and the CMS detector provided by the
following funding agencies: BMWF and FWF (Austria); FNRS and FWO (Belgium); CNPq, CAPES, FAPERJ, and FAPESP (Brazil);
MEYS (Bulgaria); CERN; CAS, MoST, and NSFC (China); COLCIENCIAS (Colombia); MSES (Croatia); RPF (Cyprus); MoER,
SF0690030s09 and ERDF (Estonia); Academy of Finland, MEC, and HIP (Finland); CEA and CNRS/IN2P3 (France); BMBF, DFG, and
HGF (Germany); GSRT (Greece); OTKA and NKTH (Hungary); DAE and DST (India); IPM (Iran); SFI (Ireland); INFN (Italy); NRF
and WCU (Republic of Korea); LAS (Lithuania); CINVESTAV, CONACYT, SEP, and UASLP-FAI (Mexico); MSI (New Zealand); PAEC
(Pakistan); MSHE and NSC (Poland); FCT (Portugal); JINR (Armenia, Belarus, Georgia, Ukraine, Uzbekistan); MON, RosAtom,
RAS and RFBR (Russia); MSTD (Serbia); SEIDI and CPAN (Spain); Swiss Funding Agencies (Switzerland); NSC (Taipei);
ThEPCenter, IPST and NSTDA (Thailand); TUBITAK and TAEK (Turkey); NASU (Ukraine); STFC (United Kingdom); DOE and NSF
(USA).

\bibliography{auto_generated}   

\cleardoublepage \appendix\section{The CMS Collaboration \label{app:collab}}\begin{sloppypar}\hyphenpenalty=5000\widowpenalty=500\clubpenalty=5000\input{EXO-12-010-authorlist.tex}\end{sloppypar}
\end{document}

%% file: EXO-12-010-authorlist.tex
\textbf{Yerevan Physics Institute,  Yerevan,  Armenia}\\*[0pt]
S.~Chatrchyan, V.~Khachatryan, A.M.~Sirunyan, A.~Tumasyan
\vskip\cmsinstskip
\textbf{Institut f\"{u}r Hochenergiephysik der OeAW,  Wien,  Austria}\\*[0pt]
W.~Adam, E.~Aguilo, T.~Bergauer, M.~Dragicevic, J.~Er\"{o}, C.~Fabjan\cmsAuthorMark{1}, M.~Friedl, R.~Fr\"{u}hwirth\cmsAuthorMark{1}, V.M.~Ghete, N.~H\"{o}rmann, J.~Hrubec, M.~Jeitler\cmsAuthorMark{1}, W.~Kiesenhofer, V.~Kn\"{u}nz, M.~Krammer\cmsAuthorMark{1}, I.~Kr\"{a}tschmer, D.~Liko, I.~Mikulec, M.~Pernicka$^{\textrm{\dag}}$, D.~Rabady\cmsAuthorMark{2}, B.~Rahbaran, C.~Rohringer, H.~Rohringer, R.~Sch\"{o}fbeck, J.~Strauss, A.~Taurok, W.~Waltenberger, C.-E.~Wulz\cmsAuthorMark{1}
% ...
\vskip\cmsinstskip
\dag:~Deceased\\
1:~~Also at Vienna University of Technology, Vienna, Austria\\
2:~~Also at CERN, European Organization for Nuclear Research, Geneva, Switzerland\\
% ...
